\documentclass{rspublic_pp}
\usepackage{amsfonts}
\usepackage{amssymb}
\usepackage[numbers,sort&compress]{natbib}
\usepackage{graphicx}
\usepackage{amsmath}
\usepackage{appendix}
\usepackage{amsthm}
\usepackage{epigraph}

\input{tcilatex}
\begin{document}

\title[Mixing, tunnelling and the direction of time]{Mixing, tunnelling and the
direction of time in the context of Reichenbach's principles}
\author[A.Y. Klimenko]{A.Y. Klimenko \thanks{%
email: klimenko@mech.uq.edu.au}}
\maketitle

\begin{abstract}
{the direction of time, the second law of thermodynamics, mixing,
decoherence, quantum tunnelling, the time primer} This work reviews the
understanding of the direction of time introduced by Hans Reichenbach,
including the fundamental relation of the perceived flow of time to the
second law of thermodynamics (i.e. the Boltzmann time hypothesis), and the
principle of parallelism of entropy increase. An example of a mixing process
with quantum effects, which is advanced here in conjunction with
Reichenbach's ideas, indicates the existence of a presently unknown
mechanism that reflects global conditions prevailing in the universe and
enacts the direction of time locally (i.e. the "time primer"). The
possibility of experimental detection of the time primer is also discussed:
if the time primer is CPT-invariant, its detection may be possible in
high-energy experiments under the current level of technology.
\end{abstract}

\affiliation{The University of Queensland, SoMME, QLD 4072, Australia} 
\epigraph{It appears that mixing processes, in the most general sense of the
term, are the instruments which indicate a direction of time}{Hans
Reichenbach}

\section{Introduction}

Discussing time is always difficult since the notion of time is deeply
embedded into both our language and our intuition. Many key words in English
(e.g. \textquotedblleft then\textquotedblright ,\textquotedblleft
follows\textquotedblright , \textquotedblleft since\textquotedblright ) and
most other languages and cultures imply both a logical link and a temporal
arrangement. The perceived flow of time and conceptual inferences are almost
indistinguishable, or at least they are not properly distinguished by most
languages we use. Immanuel Kant \cite{Kant2007} wrote in 1781:

\begin{quote}
Time is a necessary representation that grounds all intuitions. In regard to
appearances in general one cannot remove time, though one can very well take
the appearances away from time. Time is therefore given a priori.
\end{quote}

\noindent One one hand this intuition assists us in everyday life and in the
formulation of scientific theories not related to the nature of time. One
the other hand, this intuition needs to be subordinated to rational thought
when the nature of time is discussed, and this can be difficult. It is
worthwhile to note that the conventional intuitive interpretation of the
flow of time is the most common interpretation, but certainly not the only
one possible: there are indigenous tribes living in the North-Western part
of Queensland, who intuitively perceive time as being directed from East to
West.

The perceived flow of time is thought to reflect causality --- the
fundamental directional connection between events unfolding in time, as well
as the possibility of explaining observed phenomena in terms of more basic
principles. The two sides of causality, related to 1) atemporal logical
statements of a generic nature (e.g. objects fall because of the action of
gravity) and 2) directional dependence between specific consecutive events
(the vase is shattered because it was pushed from the table), may be
interpreted synergistically \cite{Bunge} or be clearly distinguished \cite%
{BornM1964}. It is the second interpretation, which is often referred to as 
\textit{antecedent causality}, that we are interested most in this work. In
the 1st half of the 20th century, there was a common belief that the
directional properties of the perceived flow of time can be explained in
terms of more objective casual relations that are postulated a priori as one
of the fundamental intrinsic properties of nature. This belief had to face
mounting difficulties in defining causality, and largely evaporated toward
the end of the 20 century. As early as in 1914, Bertrand Russell \cite%
{Russell2009} noted that

\begin{quote}
The view that the law of causality itself is a priori cannot, I think, be
maintained by anyone who realises what a complicated principle it is.
\end{quote}

The conceptual understanding of causality has grown to accommodate
randomness, counterfactual logic, etc. but, overall, our interpretation of
causality remains largely intuitive and rather short of being the basis of
rational thought. Antecedent causality is now explained in terms of physical
laws rather than placed at the foundation of these laws. Dowe \cite{Dowe1992}%
\ defines the direction of casual action in terms of physical laws that
possess temporal asymmetry: either the second law of thermodynamics or CP
violations in the quantum world. Tying causality to the second law of
thermodynamics in one form or another has become the central element of
conventional thinking about the problem (\cite%
{Time1997Faye,Albert2000,Albert2015,Loewer2007}). The strongest form of the
link between the direction of time and the second law of thermodynamics is
given by the Boltzmann time hypothesis, which proclaims that the arrow of
time and the second law are two sides of the same physical effect \cite%
{Boltzmann-book,Reichenbach1971,Hawking1993,Klimenko2019}. Hawking \cite%
{Hawking1993} explains this: \textquotedblleft the second law of
thermodynamics is really a tautology\textquotedblright , since the direction
of our perceived flow of time is, in fact, determined by the second law. The
physical side of the direction of time is covered in a number of principal
publications \cite{Prig1980,PenroseBook,Zeh2007,Hawking-time2011}.

The second half of the 20th century is marked by two seminal, yet very
different, books that endeavour to bridge philosophical and physical
arguments about the direction of time \cite{Reichenbach1971,PriceBook}. The
book by Huw Price is well written and delivers its message

\begin{quote}
I have been trying to correct a variety of common mistakes and
misconceptions about time in contemporary physics --- mistakes and
misconceptions whose origins lie in the distorting influence of our own
ordinary temporal perspective, and especially of the time asymmetry of that
perspective
\end{quote}

\noindent in a clear and articulate form. The other book is the last book
written by Hans Reichenbach. He was not able to complete his work and the
book was published by Mrs. Reichenbach in 1956, after her husband's death in
1953. The book tends to mix philosophical and physical arguments in a way
that might be confusing for both philosophers and physicists, yet
Reichenbach's book is probably the greatest book about time ever written.
According to his wife, Reichenbach considered his last book to be the
culmination of his contribution to philosophy. The \textit{Boltzmann time
hypothesis}, the \textit{principle of parallelism of entropy increase} and 
\textit{the principle of the common cause} are, perhaps, the most important
contributions presented in the book. While the Boltzmann time hypothesis
gradually became accepted by many philosophers and physicists, the principle
of parallelism of entropy increase is still a subject of debates \cite%
{Davies1977,Sklar1993,Albert2000,Albert2015,Winsberg2004b,North2011}.

The present work is, of course, not intended to review all issues related to
the arrow of time and causality within a short article. Conceptual issues
are discussed only in the context of selected examples that can illustrate
physical statements in a concise and transparent manner. Without attempting
to overview or replace the comprehensive publications cited above, this work
focuses on select few problems. Section \ref{SecBOL} briefly overviews the
understanding of the directionality of time suggested by Reichenbach.
Section \ref{SecMIX} analyses an example of a mixing process and
demonstrates the significance of time priming pointing to existence of
unknown physical mechanisms of very small magnitude associated with the
direction of time. Section \ref{SecDIS} discusses a wider scope of issues
focusing on the possibility of experimental evaluation of these mechanisms.
The Appendix considers the example of Section \ref{SecMIX} and involves
evaluation of a quantum system in thermodynamic conditions when decoherence
or recoherence are present.

\section{ The direction of time and the second law \label{SecBOL}}

Our experience of time is very directional --- we remember the past but
cannot possibly remember the future and our photographs always show us
younger than we are now. If we see dents on bumpers of two cars that are
standing next to each other, we conclude that these cars have just collided
and, certainly, not that they are going to collide in the future. At an
intuitive level, we characterise these directional properties of time as
\textquotedblleft time flow\textquotedblright\ but, according to the
fundamental Boltzmann time hypotheses, these properties of time reflect the
objective reality and directional nature of the second law of
thermodynamics. Unlike most physical theories (e.g. classical and quantum
mechanics, relativity and electromagnetism) which are time-symmetric, this
law is time-directional, stating that, in an isolated system, entropy must
increase (or stay the same) forward in time. Following Reichenbach, the
Boltzmann time hypotheses is explained below by using a gedanken experiment
called \textquotedblleft footsteps on a beach\textquotedblright .

\begin{figure}[h]
\begin{center}
\includegraphics[width=12cm,page=1,trim=0cm 5.5cm 0cm 6.8cm, clip ]{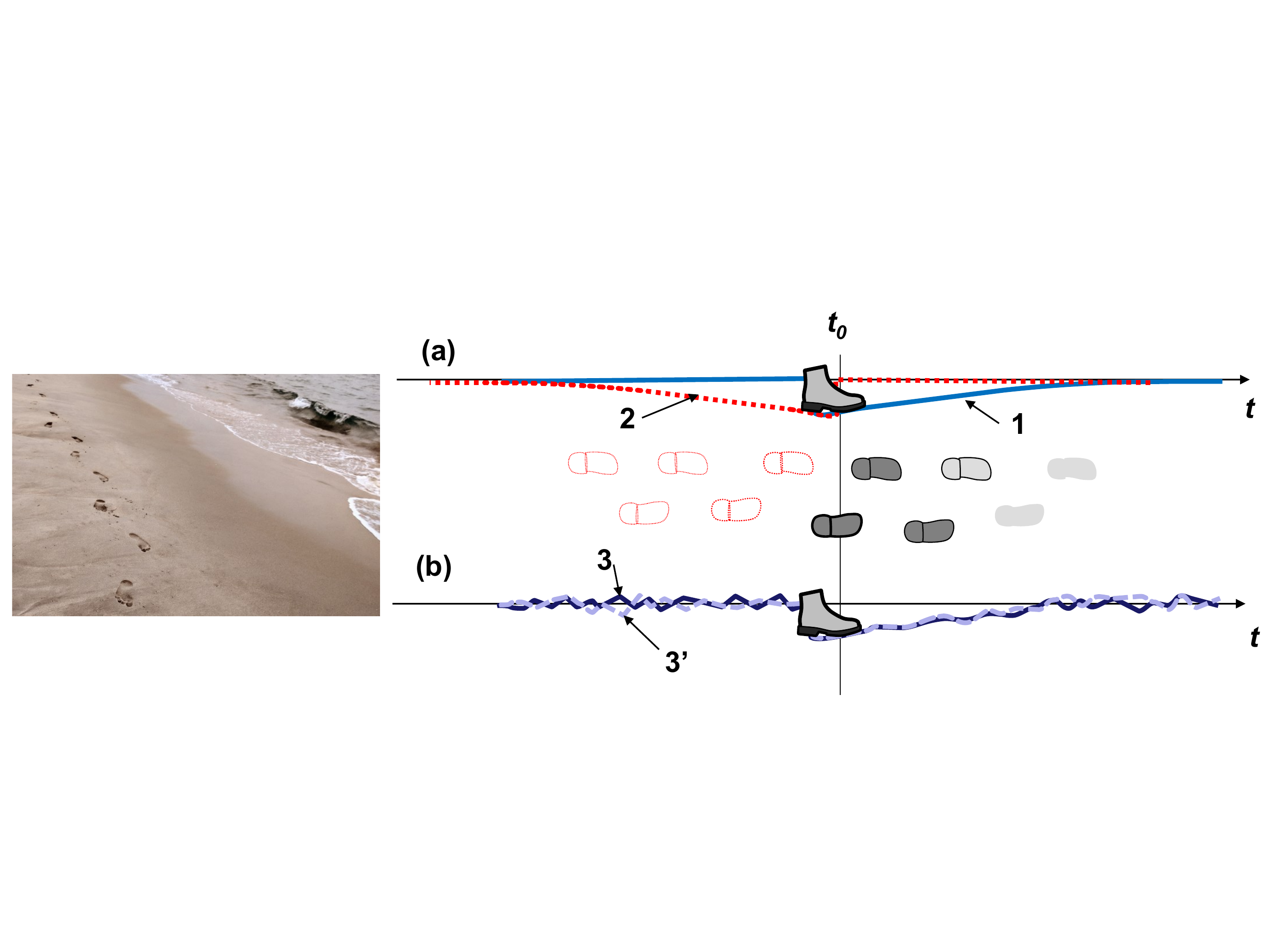}
\end{center}
\caption{Footprints on a beach: a) effect of the second law of
thermodynamics and b) effect of random disturbances. Curves: 1-realistic;
2-violating the second law; 3,3'- realistic disturbed by wind.}
\label{fig1}
\end{figure}

\subsection{Why don't we remember the future?}

The sand on a beach is always levelled by wind and water -- this is the
state of maximal entropy where all specific information is destroyed. We
might try to change this by stepping on the sand and leaving footprints.
These footprints, however, cannot stay forever and will soon disappear. This
process, shown by line 1 in Figure \ref{fig1}, is perfectly consistent with
the second law of thermodynamics. Another possibility is shown by line 2 ---
footsteps gradually appear and then are removed by a walking man. The second
scenario is not realistic as it contradicts the second law of
thermodynamics: footsteps cannot appear forward in time under the influence
of random factors such as wind and waves. To be more precise, this can, in
principle, happen, but the probability of such event is so small so that it
can be safely neglected. The second law of thermodynamics is a probabilistic
law --- it predicts the behaviour of entropy not with absolute certainty but
with overwhelming probability.

If we see footsteps on the beach, do they mean that someone walked on the
beach in the past (line 1) or that someone will walk on the beach in the
future (line 2)? According to the second law, footsteps cannot possibly
appear without a reason (i.e. a man walking) in the past but do not need a
reason to disappear. In the same way marks, photos, memories, scratches of
car paint, etc. reflect past events but tell us nothing about future events.
This conclusion is obvious but its link to the second law of thermodynamics
is not trivial.

The Boltzmann time hypothesis has not been accepted universally. Karl
Popper, one of the most distinguished philosophers of the 20th century,
argued that the Boltzmann time hypothesis cannot be true due to
thermodynamic fluctuations and that Boltzmann would not suggest his
hypothesis if he knew more about these fluctuations\ \cite{Popper1958Nature}%
. Popper's remarks are usually accurate, sharp and impressively prescient,
but this statement seems rather controversial. First, Boltzmann was well
aware of thermodynamic fluctuations and even interpreted (for the sake of
illustration) his imaginary world of reversed time as a gigantic galactic
fluctuation \cite{Boltzmann-book}. Second, exactly the same fluctuation
argument was later used not against but in support of connection between the
arrow of time and the second law of thermodynamics \cite{Dowe1992}. The flow
of time is a powerful illusion; it is very useful in real life and even in
scientific applications, but, as noted by Price \cite{PriceBook}, it can
easily produce a distorted view when issues related to the direction of time
are discussed. Although details of specific opinions may vary, most
philosophers and physicists tend to accept the existence of deep underlying
link between the perceived direction of time and the action of the second
law of thermodynamics \cite%
{Boltzmann-book,Reichenbach1971,Dowe1992,Hawking1993,PriceBook,Time1997}.

\subsection{Parallelism of entropy increase}

The importance of this principle was stressed by Reichenbach,\ who
considered the main system to be divided into branch systems (i.e
semi-independent subsystems branching from the main system) and suggested
that \textquotedblleft \textit{in the vast majority of branch systems, the
directions toward higher entropy are parallel to one another and to that of
the main system}\textquotedblright . Since \textquotedblleft the main
system\textquotedblright\ can be deemed to encompass the whole universe, its
direction toward higher entropy is the temporal direction of overall entropy
increase in the universe. This principle does not preclude occasional
fluctuations that might slightly decrease local entropy and, therefore, it
is not clear to what extent this principle represents an independent
statement. For example, Boltzmann believed that local entropy trends simply
reflect global increase of entropy in the observable part of the universe,
while Reichenbach insisted that parallelism of entropy increase is an
independent principle, which cannot be inferred from the global temporal
conditions imposed on the universe. Since a microstate of each branch system
can be characterised by a point in a phase space of very large dimensions,
the state of maximal entropy corresponds to the uniform distribution of such
points over all possible microstates. Reichenbach interprets increase of
entropy as a \textit{generalised mixing} process, which is associated with
diffusion of particles or points towards being distributed over larger
volumes in the physical and/or phase spaces. This interpretation of the
entropic directionality as a trend to expand distributions in phase spaces
of large dimensions is often used by physicists \cite{PenroseBook}. The
principle of parallelism of entropy increase is explained and critically
evaluated in a few publications, most notably in books by Davies \cite%
{Davies1977} and Sklar \cite{Sklar1993}.

While association of causality with the second law is now widely
acknowledged, the physical origins of the second law remain unclear. The
second law is fundamental but largely empirical: it declares that entropy
increases forward in time but does not explain why. Since all major physical
laws and theories are time symmetric, the most common explanation is that
the temporal asymmetry of the second law is due to asymmetric temporal
boundary conditions imposed on the universe (these conditions can be
referred to as the past hypothesis or low-entropy Big Bang). Albert \cite%
{Albert2000,Albert2015} believes that this explanation is perfectly
sufficient but, according to Reichenbach, the principle of parallelism of
entropy increase is needed (in addition to the commonly presumed low-entropy
conditions in the early universe) to explain the observed consistency of the
second law \cite{Reichenbach1971}. Winsberg \cite{Winsberg2004b} agrees with
Reichenbach, while North \cite{North2011} supports Albert. As discussed
further in Section \ref{SecMIX}, there are reasonable arguments on both
sides of this debate but, overall, it seems rather unlikely that can be
replaced by a combination time-symmetric physical laws and time-assymetric
temporal boundary conditions.

The principle of parallelism of entropy increase allows us to apply entropic
considerations to relatively small thermodynamic systems or even to
non-thermodynamic macroscopic objects. We often imply this principle when we
commingle macroscopic and microscopic considerations. For example, one can
associate an entropy change to random reshuffling of playing cards, although
this change is insignificant compared to changes in thermodynamic entropy
--- the latter is larger by a factor of $\sim 1/k_{B},$ where $k_{B}$ is the
Boltzmann constant. While applying entropic considerations to macroscopic
objects mostly produces reasonable outcomes and good intuitive illustrations
of thermodynamic principles, such applications are less rigorous compared to
the very high level of statistical certainty associated with the laws
involving the thermodynamic entropy. Macroscopic interpretations of entropy
are subject to conditions that are difficult to stipulate in a rigorous and
universal manner and, therefore, may produce incorrect inferences if taken
out of context. Reichenbach notes that we can put cards back into their
original order if we need to, but we cannot possibly reorder molecules
exactly into their original positions. The grains of sand from the example
shown in Figure \ref{fig1} may be very small, but they are macroscopic
objects. \ 

\subsection{ The principle of the common cause}

Reichenbach states this principle as\ \textquotedblleft if an improbable
coincidence has occurred, there must exist a common cause\textquotedblright
; this cause should be in the past as common effects in the future cannot
cause improbable coincidences. The term \textquotedblleft improbable
coincidence\textquotedblright\ for two events A and B, refers to the
simultaneous occurrence of A and B in excess of $P($A$)P($B$)$ --- the
probability if they were independent events. Price refers to this property
as the principle of the independence of incoming influences (PI$^{3}$) ---
indeed incoming influences (i.e. those that do not have a common cause) must
be statistically independent. This principle is intuitively obvious but,
again, the essence of the Boltzmann time hypothesis is that this effect is,
in fact, a consequence of the second law. Figure \ref{fig1}b illustrates
this point. Consider a model when wind and waves naturally impose some
degree of roughness on the sand level. The lines 3 and 3' shown in this
figure correspond to the effect of wind and waves causing the surface at two
selected points to fluctuate at random. \ These points level out only if
only someone steps on them. Levelling, however, does not last for long,
since wind and waves gradually introduce new disturbances, which erase the
footprints. The usual state of the surface is rough and influences of events
cannot propagate back in time (since this propagation specified by curve 2
contradicts the second law) --- these conditions require that dependences
are induced by past events.

It is probably true that Reichenbach's treatment of mutual causes and mutual
effects in his last book presents a combination of physical and
philosophical arguments, intermixing them to extent that may become puzzling
for both physicists \cite{Hawking1993} and philosophers \cite%
{sep-reichenbach}. Perhaps applying these ideas to conventional elements of
statistical physics can provide a more transparent illustration. In the next
subsection, we give an example of chemical kinetics that illustrates
Reichenbach's key point --- the link between the principle of the common
cause and the second law of thermodynamics.

\subsection{Chemical kinetics and causality \label{SecChem}}

Consider the following reactions 
\begin{equation}
\text{1) }A+B\longrightarrow AB,\text{ \ \ 2) }AB\longrightarrow A+B
\label{RR}
\end{equation}%
which are assumed not to have any heat effect. As illustrated in Figure \ref%
{fig2}, these reactions can be interpreted as open (left) and closed (right)
casual forks analysed by Reichenbach, who denoted AB by C or E. Events A, B,
AB respectively denote appearance of molecules $A,$ $B,$ $AB$ in a volume $%
V, $ \ which is much smaller than $V_{t}$ --- the total volume under
consideration. In the first reaction, A and B are causes that have a common
effect AB, while in the second reaction, $A$ and $B$ are effects that have a
common cause AB. Hence, according to the principle of the common cause $P($%
A+B$)=P($A$)P($B$)$ for the first reaction but not for the second. Here, $P($%
A+B$)$ is the probability of simultaneous presence of $A$ and $B$ in the
volume $V$. Considering that A and B are independent causes of the first
reaction and AB is the cause of the second reaction, the overall reaction
rates of the first and second reactions can be expressed by 
\begin{equation}
W_{1}=V_{t}K\frac{N_{A}}{V_{t}}\frac{N_{B}}{V_{t}},\text{\ }W_{2}=V_{t}K%
\frac{N_{AB}}{V_{t}}  \label{RW1}
\end{equation}%
where $P(X)=N_{X}V/$ $V_{t}$ for any $X=A,$ $B,$ $AB,$ $N_{X}$\ is the total
number of molecules $X$ in the volume $V_{t}$ and $K$ is the reaction rate
constant. Note that kinetic equation 
\begin{equation}
\frac{dN_{A}}{dt}=\frac{dN_{B}}{dt}=-\frac{dN_{AB}}{dt}=W_{2}-W_{1}
\label{RN}
\end{equation}%
implies that the entropy defined as 
\begin{equation}
S=N_{A}\ln \left( e\frac{V_{t}}{N_{A}}\right) +N_{B}\ln \left( e\frac{V_{t}}{%
N_{B}}\right) +N_{AB}\ln \left( e\frac{V_{t}}{N_{AB}}\right)  \label{RS}
\end{equation}%
cannot decrease; i.e. 
\begin{equation}
\frac{dS}{dt}=\frac{dN_{A}}{dt}\ln \left( \frac{V_{t}N_{AB}}{N_{A}N_{B}}%
\right) =K\left( N_{AB}-\frac{N_{A}N_{B}}{V_{t}}\right) \ln \left( \frac{%
V_{t}N_{AB}}{N_{A}N_{B}}\right) \geq 0  \label{RS1}
\end{equation}%
in accordance with the second law of thermodynamics.

\begin{figure}[h]
\begin{center}
\includegraphics[width=6cm,page=2,trim=7cm 2.5cm 9cm 15cm, clip ]{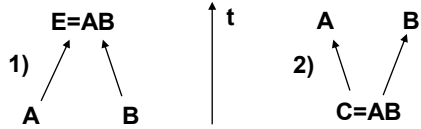}
\end{center}
\caption{Chemical reations shown in the form of casual forks.}
\label{fig2}
\end{figure}

We may try alternative anticasual arrangements when causes are located in
the future and effects are in the past. According to the anticasual
assumptions, the first reaction is caused by AB while the second reaction is
caused by two independent events A and B. This means that the overall
reaction rates are now 
\begin{equation}
W_{1}=V_{t}K\frac{N_{AB}}{V_{t}},\text{\ }W_{2}=V_{t}K\frac{N_{A}}{V_{t}}%
\frac{N_{B}}{V_{t}}  \label{RW2}
\end{equation}%
so that the entropy change rate is given by 
\begin{equation}
\frac{dS}{dt}=\frac{dN_{A}}{dt}\ln \left( \frac{V_{t}N_{AB}}{N_{A}N_{B}}%
\right) =K\left( \frac{N_{A}N_{B}}{V_{t}}-N_{AB}\right) \ln \left( \frac{%
V_{t}N_{AB}}{N_{A}N_{B}}\right) \leq 0  \label{RS2}
\end{equation}%
This illustrates that casual or anticasual assumptions imply the following
trends for the entropy: increasing in time for the former and decreasing in
time for the latter. Of course, only the casual case corresponds to the real
world.

If quantum effects are to be considered (it is arguable that interactions of
atoms are determined by quantum effects), then the casual case (\ref{RW1})-(%
\ref{RS1}) corresponds to persistent decoherence of the molecules before and
after the reaction, while the anticasual case (\ref{RW2})-(\ref{RS2})
corresponds to persistent recoherence \cite{Ent2017}. There is a physical
connection between causality and the temporal direction of decoherence \cite%
{SciRep2016,Ent2017}. The second law of thermodynamics is a macroscopic law,
but it is enacted by microscopic irreversible processes --- quantum
decoherences and collapses \cite%
{PriceBook,PenroseBook,Zeh2007,Hawking-time2011}. (We tend to use these the
terms \textquotedblleft decoherences\textquotedblright\ and
\textquotedblleft collapse\textquotedblright\ interchangeably, as there is a
significant overlap between implications of these terms --- see Appendix of
Ref. \cite{Klimenko2019} for details)

\section{Why mixing is time-directional? \label{SecMIX}}

Despite temporal symmetry of the overwhelming majority of the physical laws,
entropy tends to increase or stay the same with a high degree of certainty
for any thermodynamic system, small or large. The temporal boundary
conditions imposed on the universe (e.g. a low-entropy Big Bang) must play a
key role in this trend --- these conditions are often sufficient to explain
many effects associated with directionality of time even if physical laws
are deemed to be completely time-symmetric. Indeed, if the universe has a
very strong overall trend to increase the entropy and the universe is
divided into semi-autonomous subsystems (branches according to Reichenbach),
then increase of entropy must be more likely than decrease of entropy in
these subsystems. While the low-entropy initial conditions imposed on the
universe are important and instrumental in explaining entropy increase for
many physical phenomena, this does not mean that all observed physical
effects can be directly explained by imposing these conditions while
assuming that all physical laws are strictly time-symmetric. Therefore, the
principle of parallelism of entropy increase is indicative of some
fundamental properties of the universe that need to be understood and
examined further.

These points are illustrated here by analysing time-directional properties
of mixing. We consider diffusion of $N_{t}$\ molecules (called particles) of
a substance in a gas. The number $N_{t}$ is relatively small so that
molecules do not interact with each other; the admixture remains passive and
does not affect major thermodynamic quantities such as pressure and density,
although $N_{t}$ is large enough in absolute terms to produce reliable
statistical quantities that can be observed macroscopically as
concentrations.

\subsection{Importance of the initial conditions}

The particles (molecules) $j=1,...,$ $N_{t}$ are released at the same
location $x_{j}=x_{0}$ at $t=t_{1}$ and diffuse forward in time $t>t_{1}.$
The particle trajectories $x_{j}(t)$ represent Brownian motion, while the
average concentration of particles $f(x,t)$ satisfies the diffusion equation 
\begin{equation}
\frac{\partial f}{\partial t}=D\frac{\partial ^{2}f}{\partial x^{2}}
\label{D1}
\end{equation}%
Note that particle trajectories are time-symmetric --- that is we cannot
distinguish trajectories running forward in time from those running backward
in time. The direction of diffusion is determined by the initial conditions $%
x_{j}=x_{0}$ at $t=t_{1}.$ If we impose final conditions at $t=t_{2}>t_{1}$
\ (this can be done by considering the following process $x_{j}^{\prime
}(t)=x_{j}(t)-x_{j}(t_{2})+x_{0},$ which satisfies $x_{j}^{\prime }=x_{0}$
at $t=t_{2}),$\ then the concentration of trajectories $x_{j}^{\prime }(t)$
would be characterised by diffusion equation (\ref{D1}) but with a negative
diffusion coefficient $D^{\prime }=-D;$ i.e. this is diffusion occurring
backward in time\footnote{%
Note that this reversal is different from the reversal of the Kolmogorov
backward equation and time reversal of Markov diffusion processes preserving 
$f(x,t)$ --- see ref. \cite{K_QJMAM}. It is also possible to use both
conditions at $t=t_{1}$ and $t=t_{2}$, leading to so called Brownian bridge,
but this case is not considered here.}.

This seem to favour temporal boundary conditions as a driving force behind
irreversibility. The processes described by the diffusion equation with
positive and negative diffusion coefficients are radically different. The
direction of the diffusion is determined not by the random variations of
particle positions, which do not have a time arrow, but by imposing the
initial or the final conditions. The influence of initial or final
conditions, however, disappears in the equilibrium state $f=\func{const}$ of
fully mixed components (assuming that the diffusion takes place in a finite
volume). Indeed, once the steady-state is achieved, say within the interval $%
t_{1}^{\circ }<t<t_{2}^{\circ }$ where $t_{1}<t_{1}^{\circ }<t_{2}^{\circ
}<t_{2}$, it is impossible to tell the direction of the diffusion process,
even if the most detailed current characteristics of trajectories are
monitored --- information about initial or final conditions has been lost.
Setting initial conditions at $t=t_{1}$ cannot be distinguished from setting
the final conditions at $t=t_{2}$ by observing equilibrium solution at $%
t_{1}^{\circ }<t<t_{2}^{\circ }$. Equilibria achieve maximal entropy and
destroy information.

The evolution of the universe can be interpreted as a generalised mixing
process where particles diffuse to occupy a larger and larger number of
microstates. Since the universe was presumably formed with low-entropy
initial conditions and has not achieved its equilibrium state, this
consideration provides a justification for generally preferring initial
conditions to final conditions in today's environment. It might seem that
Reichenbach's\ principle of parallelism of entropy increase is excessive ---
the low-entropy initial condition imposed on the universe ensures both
overall entropy increase and, as long as overall equilibrium is not reached,
proper directionality of various local thermodynamical processes. While on
many occasions global entropy increase would enforce entropy increases in
local processes, there is an important detail that is missing in this
inference.

\begin{figure}[h]
\begin{center}
\includegraphics[width=12cm,page=3,trim=1.5cm 1cm 0.2cm 3.5cm, clip ]{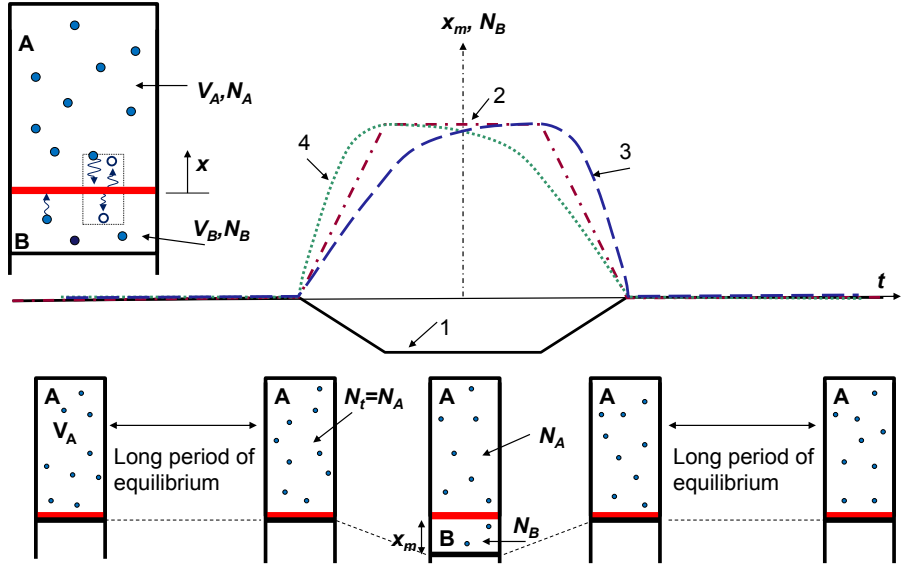}
\end{center}
\caption{Experiment with admixture passing through semi-permeable membrane.
Curves: 1 -- position of the piston $x_m(t)$; 2 -- equilibrium $N_B(t)$; 3
-- $N_B(t)$ for $C=+1$; 4 -- $N_B(t)$ for $C=-1$. }
\label{fig3}
\end{figure}

\subsection{Why is the principle of parallelism of entropy increase
essential? \label{SecPar}}

At this point we consider a modified experiment, which is illustrated in
Figure \ref{fig3}. A cylinder having a finite volume $V_{t}$ contains $N_{t}$
particles of the passive admixture (as considered previously) and is kept in
a state of thermodynamic equilibrium for a long time. The cylinder is
located in a remote part of the universe away from any possible influences
of the matter that populates the universe. The piston remains at $x=0$ for a
long time so that $V_{\text{A}}=V_{t}$ and $V_{\text{B}}=0$, then moves down
and up in a time-symmetric manner so that $V_{\text{B}}>0$ as shown in
Figure \ref{fig3} and then, again, remains at $x=0$ for a very long time so
that $V_{\text{B}}=0.$ In addition to admixture molecules, the cylinder may
also be filled by a gas to ensure that the system under consideration is
thermodynamic. The volumes A and B\ are divided by a very thin
semi-permeable membrane that is fully permeable for the gas (if gas is
present) and only partially permeable for the molecules of admixture, so
that these molecules can occasionally tunnel through the membrane. When
considered from a quantum-mechanical perspective, the membrane is
interpreted as a potential barrier that can be tunneled through, while the
rest of the walls are formed by impervious barriers of a high potential. We
note that such experiments are not only conceptually possible but, due to
recent technological advances \cite{QTG2018}, also practically feasible.
Obviously, $N_{\text{A}}+N_{\text{B}}=$ $N_{t}=\func{const}$ and $V_{\text{A}%
}=\func{const}.$ The number of particles $N_{t}$ is sufficiently large to
ensure that $N_{\text{A}}$ and $N_{\text{B}}$ are macroscopic parameters,
which can be measured by classical instruments.

For simplicity of evaluation, the probability of successful tunnelling of
admixture molecules through the membrane is assumed to be small so that the
concentrations of particles remain uniform in volumes A and B (although not
necessarily the same on both sides of the membrane --- see Figure \ref{fig3}%
). Since particles do not interact, they can be considered autonomously from
one another. The concentrations of particles on both sides of the membrane
are determined by quantum tunnelling through the membrane. Classical
statistics is assumed so that most of the quantum states are vacant: all
states have the same probability of occupation determined by the
concentrations of the particles: $N_{\text{A}}/V_{\text{A}}$ on one side and 
$N_{\text{B}}/V_{\text{B}}$ on the other.

As the particles tunnel through the membrane, they must decohere since,
otherwise they would be simultaneously present in volumes A and B, be
governed by unitary evolutions and not subject to the laws of statistical
physics (see Ref. \cite{matrix2019}). We, however, do not have any
experimental evidence that this can happen when an object is progressively
screened from the direct influence of the initial and final conditions
imposed on the universe. If decoherence is terminated, we would effectively
obtain a less cruel version of Schr\"{o}dinger's cat --- a substance whose
particles are not located in volumes A or B but are in superposition states
between these volumes (strictly speaking, $N_{\text{A}}$ and $N_{\text{B}}$
are not classically defined in this case). After decoherence and collapse of
the wave function, particles appear either on one side of the membrane or
the other with some classical probability.\ As we do not wish to
discriminate the direction of time a priori, we must admit that the
particles can decohere or recohere (i.e. decohere backward in time), as
discussed in the Appendix. The concentration of particles is governed by the
equation (see Appendix and Refs. \cite{SciRep2016,Ent2017})%
\begin{equation}
\frac{dN_{\text{B}}}{dt}=-\frac{dN_{\text{A}}}{dt}=CK\left( \frac{N_{\text{A}%
}}{V_{\text{A}}}-\frac{N_{\text{B}}}{V_{\text{B}}}\right)  \label{Eq1}
\end{equation}%
where $K$ is the rate constant for transition through the membrane, which,
as shown in the Appendix, must be the same for transitions from A to B and
from B to A. The constant $C=+1$ corresponds to predominant decoherence and $%
C=-1$ to predominant recoherence (i.e. decoherence back in time). In
principle, we also need to consider the case of $C=0$ (assuming that
intensities of decoherence and recoherence exactly match each other) but
this case is not realistic. Indeed, if the piston moves very slowly, then
the densities of particles must approach the same values on both sides of
the membrane\ $N_{\text{A}}/V_{\text{A}}=N_{\text{B}}/V_{\text{B}}=$ $N_{t}/$
$V_{t}$\ and, obviously, $N_{\text{A}}(t)=V_{\text{A}}N_{t}/$ $V_{t}(t)$. On
the one hand, $N_{\text{A}}(t)$ changes but, on the other hand, equation (%
\ref{Eq1}) with $C=0$ enforces that $dN_{\text{A}}/dt=0$. Therefore,
particles must either predominately decohere or predominantly recohere. This
can be easily determined by moving the piston a bit faster so that the
solution of equation (\ref{Eq1}) deviates from the equilibrium given by $N_{%
\text{A}}(t)=V_{\text{A}}N_{t}/$ $V_{t}(t)$, as illustrated in Figure \ref%
{fig3}. We can observe either the behaviour indicated by line 3, which
corresponds to $C=+1$, or the behaviour indicated by line 4, which
corresponds to $C=-1.$ The difference between the two cases is in the
definition of the direction of time. As we use the conventional definition
of the direction of time, where entropy increases toward $t=+\infty $ , then 
$C=+1$ and particles predominantly decohere.

From the perspective of quantum mechanics, the state of equilibrium
corresponds to the maximally mixed quantum state, where the density matrix
is proportional to the unit matrix and the entropy is maximal. This state of
maximal entropy cannot be altered without external interference; neither by
unitary evolution (which cannot change entropy), nor by decoherence (which
cannot reduce entropy). The effect of decoherence, therefore, is not
observable in equilibrium conditions (as it should be --- equilibrium states
do not evolve). It would be rather unphysical to assume that decoherence,
which exists at its full strength under smallest deviations from
equilibrium, physically disappears once full equilibrium state is reached.
It is the statistical effect of decoherence that disappears, not decoherence
itself: it still affects individual particles at microscopic level. This can
be illustrated by the Ehrenfest urn model: balls are located in two urns are
picked up at random and are placed into another urn (possibly with a fixed
probability reflecting the transmission rate between the urns). Each act of
redistribution of balls increases uncertainty of ball locations, and
therefore, increases the corresponding entropy. Once equilibrium is reached
and the two urns have the same number of balls, the process (which still
continues physically) does not change the distribution and does not change
the entropy.

We observe a very interesting situation: the system stays in complete
equilibrium for a very long time and should not be affected by any initial
conditions that were imposed on the system or on the whole universe a long
time before the experiment. According to the conditions of the experiment,
all external influences must be macroscopic. These influences are limited to
the movements of the piston, which are conducted in a time-symmetric manner
and cannot possibly create any directionality of time. The known laws of
classical, quantum and relativistic physics are also time-symmetric. Why do
the particles behave in a time-directional manner (decohere and not
recohere)? Reichenbach's principle of parallelism of entropy increase
clearly requires that $C=+1$ in (\ref{Eq1}) and, at least under conditions
shown in Figure \ref{fig3}, this cannot be directly explained by the
low-entropy initial conditions imposed on the universe. Something must be
missing.

\subsection{The time primer}

We, of course, do not suggest that predominance of decoherence ($C=+1$ in (%
\ref{Eq1})) is not related to the low-entropy initial conditions imposed on
the universe, but rather observe that there must be a physical mechanism
that connects decohering properties of matter to the fundamental state of
the universe. There is, however, no obvious or known mechanism that
translates a low-entropy Big Bang into the fact that matter predominantly
decoheres under conditions when matter is screened from the Big Bang by an
equilibrium state (presumably destroying all information about the previous
states of the universe). We can call this mechanism the \textquotedblleft
time primer\textquotedblright \cite{Klimenko2019}. The time primer is
related to the most fundamental properties of matter and its primary effect
should be predominance of quantum decoherence, resulting in the second law
of thermodynamics, causality and in the perceived \textquotedblleft flow of
time\textquotedblright . The time primer must exist and, at least in
principle, should be represented by a mechanism that can be detected in
experiments but, as discussed in the rest of this paper, this is likely to
be a very difficult task.

\begin{figure}[h]
\begin{center}
\includegraphics[width=12.5cm,page=9,trim=0cm 0cm 0cm 0cm, clip ]{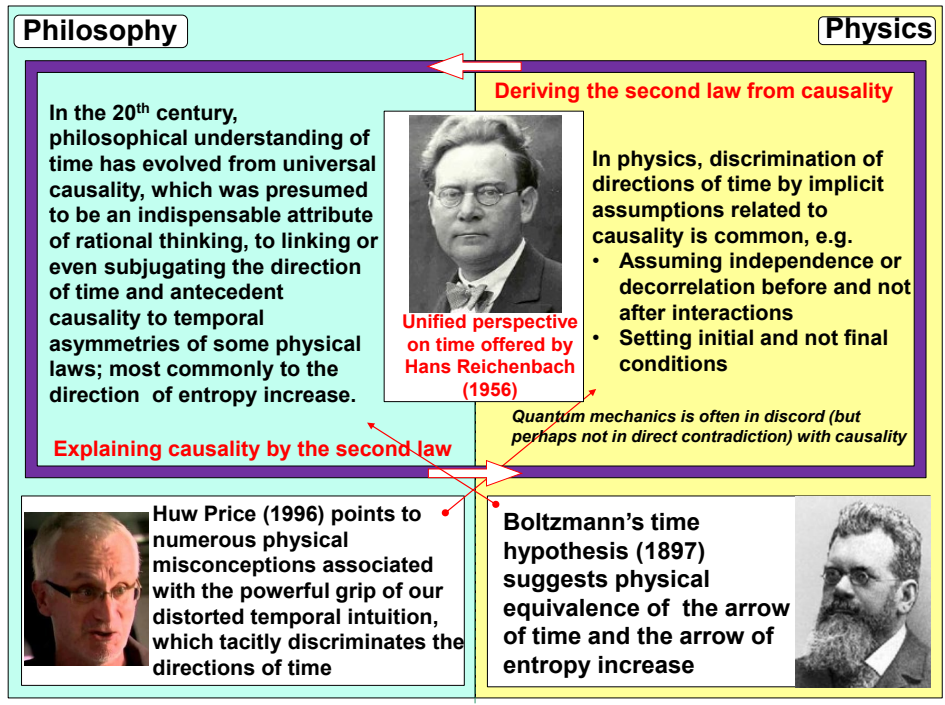}
\end{center}
\caption{If considered from a transdisciplinary perspective, arguments
commonly used in physics and philosophy in explaining antecedent causality
and the second law of thermodynamics form a logical circle }
\label{fig6}
\end{figure}

The conventional quantum theories \cite{Zurek2003,CT-P2009,Yukalov2012}
explain the physical mechanism of decoherence quite well, but only under
conditions, in which the direction of time is discriminated by implied
causality: setting initial (and not final) conditions is essential for these
theories. Therefore, we are trapped in a logical loop: we explain causality
by the second law, the second law by decoherence, and decoherence by
causality (Figure \ref{fig6}). The time primer points to an unknown physical
effect that is needed to break this loop. For the case illustrated in Figure %
\ref{fig3} there is no obvious justification for discriminating the
directions of time by preferring the initial conditions to the final
conditions. Price \cite{PriceBook} noted that physical theories often
discriminate the directions of time by intuitively implying time-directional
causality --- these may be valuable theories in many respects but they
cannot serve, as physical explanations of the directional properties of time
as these properties are presumed and not deduced.

\section{Discussion \label{SecDIS}}

The current state of arguments about direction of time (illustrated in a
simplified form by Figure \ref{fig6}) reflects persisting confusion:
philosophers seek the assistance of the physical laws (and especially that
of the second law of thermodynamics) in defining antecedent causality, while
physicists base their justifications of physical laws and theories on
implications of causality (often tacitly or implicitly). This state forms an
unsatisfactory explanatory loop, in which antecedent causality is associated
with the action of the second law and the second law is explained by the
effects of antecedent causality. While the action of the second law can be
related at an elementary level to implications of quantum decoherence and
collapse, the quantum theory, as was remarked by Einstein half-a-century
ago, still cannot provide a unified picture of physical reality. The time
primer is not a physical theory but rather a placeholder for such a theory,
recognising that something important is missing in our understanding of
thermodynamic time.

Over the last few decades, the direction of time has experienced a gradual
drift from the domain of philosophy to the domain of physics. While the
influence of physical ideas and theories gradually increases, this
transition has not been completed yet since the possibility of experimental
validation is a necessary attribute of any physical theory. The possibility
of experimental testing of the time priming is discussed in this section.

\subsection{Environmental time priming}

Decoherence may be induced by relatively weak interactions with the rest of
the universe, since our universe is far away from equilibrium and (at least
in principle) can induce time-directional effects in the system. Numerous
quantum\ theories point to environmental interferences as the mechanism
responsible for decoherence and thermodynamic behaviour in quantum systems 
\cite{Stamp2012,Zurek1982,Joos2003,EnvDec2005,CT-G2006,CT-P2006,Yukalov2012}%
. These theories, however, are not specific with respect to the physical
mechanism of interactions, which make experimental validation of these
interactions rather difficult and uncertain. Environmental interference of
CP-violating and CPT-invariant quantum systems is expected to produce
apparent CPT violations \cite{K-PhysA,Klimenko2019} and detected CPT
discrepancies \cite{Barbar2016} may be related to interference from the
environment. The problem with experimental validation of environmental
interference is that, even if this interference is detected, there is no
guarantee that it is this interference and not something else that
represents the principal mechanism controlling the time priming. To prove
this point we need to reduce this interference and expect a corresponding
reduction in consistency of time priming.

Radiation is likely to be the first suspect for thermodynamic interactions.
Since radiation itself must be decoherence-neutral \cite{Ent2017}, its role
should be in connecting the equilibrated system (in Figure \ref{fig3}) to
matter that populates the universe and remains far from equilibrium. If the
experiment is located in a remote area of the universe, incoming radiation
can be interpreted as a random signal. This signal can stimulate
decoherence, but it seems that presuming causality is unavoidable under
these conditions \cite{Zurek1982,CT-P2009,Yukalov2012}.

If a system is placed far away from all other matter, a reduction in
effectiveness of interactions can be expected. At present, however, we do
not have any evidence that thermodynamic time slows down when a system is
screened from the influence of (or placed far from) other thermodynamic
systems. Would radioactive decays become any slower if an radioactive object
is placed in a very remote part of the universe? There is no direct evidence
that this would be the case. Reichenbach believed that complete insulation
of a subsystem would not affect the rate of its entropy increase. This does
not rule out environmental mechanisms of time priming, but it does
illustrate that obtaining experimental proof of environmental time priming
would be very difficult. In principle, there might be a \textquotedblleft
time field\textquotedblright\ that is present everywhere, and the direction
of time is determined by very weak, yet very important, interactions with
this field. This case, however, is practically indistinguishable from
intrinsic mechanisms of decoherence.

\subsection{Intrinsic mechanisms of time priming}

Various theories modifying equations of quantum mechanics to incorporate
quantum collapses and decoherences have been suggested \cite%
{Penrose1996,Zurek2002, QTreview,Beretta2005,Stamp2012}. These theories,
however, assume causality rather than attempt to explain causality (and some
are merely empirical). The physical mechanism of entropy-increasing
processes at microscopic level remains uncertain. Penrose \cite%
{Penrose1996,Penrose2014a} suggested a physical mechanism that can
\textquotedblleft prime\textquotedblright\ the direction of time. This
theory (due to Diosi and Penrose) points to gravitational effects as a
culprit of irreversibilities observed in the quantum world. Gravity induces
quantum violations causing collapses of otherwise reversible unitary
evolutions. This provides a very good illustration of how small these
violations might be and how difficult it would be to directly detect them in
experiments \cite{Penrose2014a}.

Considering that radiation is expected to remain decoherence-neutral we
might extend this inference to all bosons and expect that the intrinsic
source of decoherence must be hidden in the properties of matter, most
likely in quark - containing particles (i.e. neutrons or protons) since
quarks are known of being capable to violate time symmetry (due to CP
violations). If this is the case, there remain two possibilities: baryons
and antibaryons can violate unitarity of quantum evolutions in a symmetric
or antisymmetric manner, which result in either symmetric or antisymmetric
extension of thermodynamics from matter into antimatter \cite%
{KM-Entropy2014,SciRep2016}. Symmetric and antisymmetric versions of
thermodynamics respectively correspond to CP- and CPT-invariant time priming
and can not be valid simultaneously --- only one of them can be (and is)
real. The antisymmetric version may or may not correspond to the real world
but, conceptually, it is quite attractive due to a number of reasons. One of
these reasons is that, if antisymmetric thermodynamics is valid, it
kinetically favours conversion of antimatter into matter and, at the same
time, explains the present arrow of time by the relative abundance of matter
over antimatter \cite{KM-Entropy2014,SciRep2016}. If detected in
experiments, antisymmetric thermodynamics can pinpoint at the intrinsic
mechanisms of time priming. If it is the symmetric version that is real,
then experimental examination of the intrinsic mechanisms of time priming
becomes a very difficult task.

\subsection{Testing the symmetry of time priming.}

From a theoretical perspective, testing whether thermodynamics possesses
symmetric or antisymmetric properties may seem straightforward --- we just
need to create thermodynamically significant quantities of antimatter and
see which thermodynamic properties it has. Practically, producing
significant quantities of antimatter can be extremely difficult. It might be
possible, however, to test the symmetric/antisymmetric properties of
thermodynamics at the present level of technology.

It seems that a system with some thermodynamic properties (i.e. quark-gluon
plasma \cite{Nature2007}) can be created at very small scales as a result of
collision of high-energy protons and nuclei. For example, two protons may
collide elastically producing two protons with different momenta or
inelastically producing jets of multiple particles. While the former
collisions are unitary, we are tempted to assume that the latter collisions\
have some thermodynamic features. If this thermodynamic interpretation of
inelastic collisions is correct, collisions of two antiprotons should be the
same as collisions of protons according to symmetric thermodynamics, and can
be expected to be different from collisions of protons according to
antisymmetric thermodynamics. While\ the overall energy, momentum and other
conserved properties must always be preserved, antisymmetric thermodynamics
involves opposite entropy trends for matter and antimatter. Therefore,
assuming that thermodynamic effects can play a role within very short times
associated with collisions (which is a big assumption, of course),
antisymmetric thermodynamics predicts that antiproton-antiproton collisions
should tend to have smaller inelastic collision cross-sections than the
inelastic cross-sections of the proton-proton collisions under the same
conditions. In simple terms, collisions of antiprotons should be biased
towards elastic collisions compared to collisions of protons under the same
conditions. This attributes the action of the time primer to complex
interactions of partons inside baryons, which are clearly revealed only when
collision energies are sufficiently high. The extent of the differences
between baryons and antibaryons is determined by persistency of the time
primer (i.e. it might be difficult to collide two antiprotons
inelastically). Symmetric thermodynamics does not predict any differences
between inelastic cross-sections of protons and antiprotons.

Note that the implications of antisymmetric thermodynamics may produce an
impression of CPT violations: protons and antiprotons can have different
inelastic collision cross-sections \cite{Symmetry1972}. According to
interpretation given above, this conclusion would be incorrect ---
antisymmetric thermodynamics is based on complete CPT symmetry exhibited
both at small and large scales. This effect is similar to apparent CPT
violations that can be observed due presence of the environmental mechanisms
of time priming --- see Ref \cite{K-PhysA} for details. It seems that
microscopic action of time priming can be detected (due to its interference
with unitarity) as apparently present CPT violations in systems that in fact
strictly preserve the CPT symmetry.

Another possibility for testing the extension of thermodynamics from matter
to antimatter is investigation of photon absorption and radiation by atoms
and antiatoms under the same conditions. The antiatoms need to be trapped
and cooled down, which is not easy but still possible \cite{H2anti2010}. The
kinetics of light absorption and radiation is the same for atoms and
antiatoms in symmetric thermodynamics and different in antisymmetric
thermodynamics\cite{Ent2017}. In simple terms, if antiatoms are somewhat
more reluctant to adsorb photons than the corresponding atoms under the same
conditions, then this would indicate validity of antisymmetric
thermodynamics. Again, if such effects are detected, they must not be
confused with CPT violations --- antisymmetric thermodynamics is very much
consistent with the CPT invariance.

\section{Conclusions}

This work briefly reviews and explains the principal ideas about time that
were brought by the late Hans Reichenbach in his last book. The Boltzmann
time hypothesis and the Reichenbach principle of parallelism of entropy
increase seem to be most important among these ideas. While the Boltzmann
time hypothesis tends to be accepted by modern philosophers and physicists
(at least by those who have thought about or investigated these issues), the
principle of parallelism of entropy increase is still subject to debate. In
the present work, we consider a mixing process involving quantum effects and
demonstrate that, although the low-entropy initial conditions that
characterised early universe are most important, there should be an unknown
mechanism that delivers the influence of these initial conditions to
thermodynamic subsystems observed in the real world. We call this mechanism
the \textquotedblleft time primer\textquotedblright . The time primer is
responsible for prevailing forward-time decoherence in quantum systems,
which increases entropy and, according to the Boltzmann time hypothesis,
introduces antecedent causality and other components of the perceived flow
of time.

The possibility of experimental detection of the time primer is discussed in
the last section --- in general, this task is quite difficult. If, however,
the time primer is CPT-invariant (rather than CP-invariant) and objects with
some thermodynamic properties emerge at small scales in inelastic
high-energy collisions, the direct effects of the time primer may be
detected under the current level of technology.

\bibliographystyle{unsrtnat}
\bibliography{Law3}

\appendix{Quantum tunnelling and decoherence}

\begin{figure}[h]
\begin{center}
\includegraphics[width=12cm,page=4,trim=1.2cm 3cm 2.2cm 7cm, clip ]{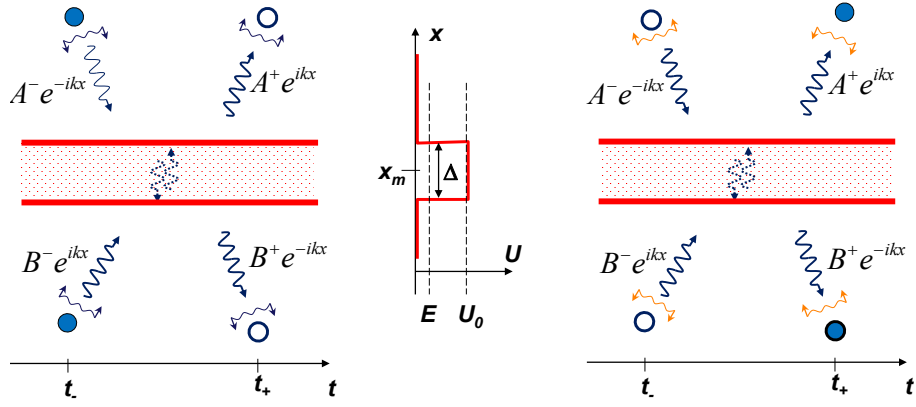}
\end{center}
\caption{Tunnelling of a particle through the membrane: left -- with
decoherence, right -- with recoherence, middle -- the membrane potential $%
U=U(x)$}
\label{fig4}
\end{figure}

The quantum outcomes of tunnelling can be expressed by the scattering matrix 
$\mathbb{S}$, which is a unitary matrix $\mathbb{SS}^{\dag }\mathbb{=I}$
that connects the amplitudes $A^{-}$ and\ $B^{-}$ of incoming waves with the
amplitudes of the outgoing waves $A^{+}$ and\ $B^{+}$ (see Figure \ref{fig4}%
) so that:%
\begin{equation}
\underset{\psi (t_{+})}{\underbrace{\left[ 
\begin{array}{c}
\tilde{A}^{+} \\ 
\tilde{B}^{+}%
\end{array}%
\right] }}=\underset{\mathbb{S}}{\underbrace{\left[ 
\begin{array}{cc}
r & q \\ 
q & r%
\end{array}%
\right] }}\underset{\psi (t_{-})}{\underbrace{\left[ 
\begin{array}{c}
\tilde{A}^{-} \\ 
\tilde{B}^{-}%
\end{array}%
\right] }},\ \ \ \underset{\psi (t_{-})}{\underbrace{\left[ 
\begin{array}{c}
\tilde{A}^{-} \\ 
\tilde{B}^{-}%
\end{array}%
\right] }}=\underset{\mathbb{S}^{\dag }}{\underbrace{\left[ 
\begin{array}{cc}
r^{\ast } & q^{\ast } \\ 
q^{\ast } & r^{\ast }%
\end{array}%
\right] }}\underset{\psi (t_{+})}{\underbrace{\left[ 
\begin{array}{c}
\tilde{A}^{+} \\ 
\tilde{B}^{+}%
\end{array}%
\right] }}  \label{Scatter}
\end{equation}%
where $\tilde{A}^{+}=A^{+}e^{ik\Delta /2},$ $\tilde{A}^{-}=A^{-}e^{-ik\Delta
/2},$ $\tilde{B}^{+}=B^{+}e^{ik\Delta /2},$ $\tilde{B}^{-}=B^{-}e^{-ik\Delta
/2}$ are the corresponding wave amplitudes evaluated at the boundaries of
the barrier at $x=\pm \Delta /2$, the asterisk denotes complex conjugates
and the values of $q$ and $r$ are specified below. The quantum barrier is
assumed to be symmetric, which corresponds to a symmetric matrix $\mathbb{S}$%
. The matrix $\mathbb{S}$ should not be confused with the commonly used
transfer matrix that links the wave amplitudes on one side of the barrier to
the wave amplitudes on the other side. Note that $\left\vert q\right\vert
^{2}+\left\vert r\right\vert ^{2}=1$ and $\left\vert r^{2}-q^{2}\right\vert
=1$ due to the unitary of $\mathbb{S}$. The tunnelling parameters $q$ and $r$
can be determined for specific shape of the potential barrier $U(x)$, which
is assumed to have a rectangular shape as shown in Figure \ref{fig4}. The
solution of this problem can be found in standard textbooks \cite{LL3}: 
\begin{eqnarray*}
r &=&(k^{2}+\kappa ^{2})\frac{(1-Q^{2})}{W},\ q=4ik\kappa \frac{Q}{W},\
Q=\exp \left( -\kappa \Delta \right) \\
W &=&(k+i\kappa )^{2}-(k-i\kappa )^{2}Q^{2},\ \ k=\frac{\sqrt{2mE}}{\hbar }%
,\ \kappa =\frac{\sqrt{2m\left( U_{0}-E\right) }}{\hbar }
\end{eqnarray*}%
\begin{equation}
\left\vert q\right\vert ^{-2}=1+\frac{1}{4}\frac{(k^{2}+\kappa ^{2})^{2}}{%
k^{2}\kappa ^{2}}\sinh ^{2}\left( \kappa \Delta \right) \underset{U_{0}\gg E}%
{\approx }\frac{1}{4}\frac{U_{0}}{E}\sinh ^{2}\left( \Delta \frac{\sqrt{%
2mU_{0}}}{\hbar }\right)
\end{equation}%
where $E$ is the energy of the particle, $\hbar $ is the Planck constant and 
$\left\vert q\right\vert ^{2}$ is the transmission coefficient. The barrier
is assumed to be thin: i.e. its thickness $\Delta $ is small but its
magnitude $U_{0}$ is large. We can assume that $U_{0}\gg E$ and therefore $%
q\ll 1$, $r\sim 1$.

The quantum description of tunnelling specified by (\ref{Scatter}) is
time-symmetric, while its effect on the thermodynamic system considered here
(Figure \ref{fig3}) is determined by the decoherence of quantum waves as
shown in Figure \ref{fig4}. The decoherence transforms the time-reversible
Schrodinger equation into the Pauli master equation, which is the principal
equation that combines quantum description with directionality of time \cite%
{Pauli1928}. The Pauli master equations are general equations that
incorporate decoherence, which determines the direction of the entropy
increase, into the quantum world; i.e. different forms of the Pauli master
equation are obtained for the same quantum system depending on properties of
decoherence and recoherence \cite{SciRep2016,Ent2017}.

Since particles do not interact and classical statistics is implied (i.e.
most quantum states are not occupied), one can consider the wave function $%
\psi _{j}$ of a single particle. The Pauli master equation for the
probabilities $p_{j}=\psi _{j}\psi _{j}^{\ast }$ (no summation over $j$) is
given by \cite{SciRep2016} 
\begin{equation}
\frac{dp_{j}}{dt}=\sum_{k}Cw_{j}^{k}p_{k}-\sum_{k}Cw_{k}^{j}p_{j}
\label{MEcpt2}
\end{equation}%
where $C=+1$ corresponds to dominant decoherence and $C=-1$ corresponds to
dominant recoherence, and $w_{j}^{k}=w_{k}^{j}$ are transitional
probabilities. Note that, unlike in Ref. \cite{SciRep2016,Ent2017}, the
predominant direction of the time priming is assumed to be the same for all
quantum states. Consider states $a=a_{1},a_{2},...$ on side A of the
membrane and states $b=b_{1},b_{2},...$ on side B of the membrane so that \ $%
j=a_{1},a_{2},...b_{1},b_{2},...$\ \ and the states $a_{i}$ and $b_{i},$ $%
i=1,2,3,...$ correspond to interacting waves with the same energy $E_{i}$.
Evaluation of the two sums over $j=a_{1},a_{2},...$ and over $%
j=b_{1},b_{2},...$ in equation (\ref{MEcpt2}) while taking into account 
\begin{equation}
\sum_{a}p_{a}=\frac{N_{\text{A}}}{N_{t}},\ \ \ \ \sum_{b}p_{b}=\frac{N_{%
\text{B}}}{N_{t}}\ ,\ \ \ N_{t}=N_{\text{A}}+N_{\text{B}}
\end{equation}%
yields 
\begin{equation}
\frac{1}{N_{t}}\frac{dN_{\text{B}}}{dt}=-\frac{1}{N_{t}}\frac{dN_{\text{A}}}{%
dt}=C\sum_{b}\sum_{a}\left( w_{b}^{a}p_{a}-w_{a}^{b}p_{b}\right)
\end{equation}%
Substituting the equilibrium distribution $g_{j}^{\circ }$ (which are
assumed to be classical Gibbs distributions due to $g_{j}^{\circ }\ll 1$)
and the density of quantum states $\rho _{j}$ 
\begin{equation}
p_{a}=\frac{N_{\text{A}}}{N_{t}V_{\text{A}}}\rho _{a}g_{a}^{\circ },\ \ \ \
p_{b}=\frac{N_{\text{B}}}{N_{t}V_{\text{B}}}\rho _{b}g_{b}^{\circ },\ \ \
g_{a_{i}}^{\circ }=g_{b_{i}}^{\circ }=\exp \left( \frac{\mu -E_{i}}{k_{B}T}%
\right)
\end{equation}%
where $\mu $ is the chemical potential, we obtain 
\begin{equation}
\frac{dN_{\text{B}}}{dt}=-\frac{dN_{\text{A}}}{dt}=C\left( K_{1}\frac{N_{%
\text{A}}}{V_{\text{A}}}-K_{2}\frac{N_{\text{B}}}{V_{\text{B}}}\right)
\end{equation}%
and%
\begin{equation}
K_{1}=\sum_{b}\sum_{a}w_{b}^{a}\rho _{a}g_{a}^{\circ
}=\sum_{i}w_{b_{i}}^{a_{i}}\rho _{a_{i}}g_{a_{i}}^{\circ },\ \ \
K_{2}=\sum_{b}\sum_{a}w_{a}^{b}\rho _{b}g_{b}^{\circ
}=\sum_{i}w_{a_{i}}^{b_{i}}\rho _{b_{i}}g_{b_{i}}^{\circ }  \label{K1K2}
\end{equation}%
since only the corresponding states $a_{i}$ and $b_{i}$ interact, i.e. $%
w_{a_{j}}^{b_{i}}=w_{b_{j}}^{a_{i}}=0$ for $j\neq i.$ The symmetry of the
coefficients $w_{a}^{b}=w_{b}^{a}$ and equilibrium distributions $%
g_{a_{i}}^{\circ }=g_{b_{i}}^{\circ }=g_{i}^{\circ }$ and the same
conditions on both sides of the membrane $\rho _{a_{i}}=\rho _{b_{i}}=\rho
_{i}$ yield equation (\ref{Eq1}) with $K=K_{1}=K_{2}$. The direction of
thermodynamic time in this equation is determined by the temporal direction
of decoherence (i.e. by $C=+1$ or $C=-1$). As expected \cite{Ent2017}, 
\textit{the transmission rate is proportional to the concentration of
decohered particles and does not depend on the concentration of recohered
particles, irrespective of the temporal direction of decoherence or
recoherence}.

Considering that the scattering matrix is close to unity, one can write $%
\mathbb{S}=\mathbb{I}+i\mathbb{T}$ where $\mathbb{T}$ is small (since $%
\left\vert q\right\vert ^{2}\ll 1$) and Hermitian $\mathbb{T}^{\dag }=%
\mathbb{T}$ at the leading order. The operator $\mathbb{T}$ can be
conventionally be expressed in terms of the interaction Hamiltonian by using
perturbation methods, but this is not needed here as we already have the
exact solution for the tunnelling problem. Substituting $%
w_{a_{i}}^{b_{i}}=w_{b_{i}}^{a_{i}}=\mathfrak{A}u_{i}\left\vert q\right\vert
_{i}^{2}/2,$ where $\mathfrak{A}$ is the area of the membrane, $\left\vert
q\right\vert _{i}^{2}\approx 4E_{i}e^{-2\kappa \Delta }/U_{0}$\ is the
transmission coefficient, $u_{i}^{2}=2E_{i}/m$ and $\kappa =\left(
2mU_{0}\right) ^{1/2}/\hbar $, into (\ref{K1K2}) results in 
\begin{equation}
K=2^{2\frac{1}{2}}\frac{\mathfrak{A}}{U_{0}m^{%
{\frac12}%
}}e^{-2\kappa \Delta }\sum_{i}E_{i}^{1\frac{1}{2}}\rho _{i}g_{i}^{\circ }
\end{equation}

A more detailed analysis of tunnelling without decoherence under these
conditions can be found in Ref. \cite{matrix2019}.

\end{document}